\documentclass[%
 reprint,
 amsmath,
 amssymb,
 aps,
prd,
twocolumn,
superscriptaddress,
preprintnumbers,
floatfix,
nofootinbib
]{revtex4-2}

\usepackage{graphicx}%
\usepackage{dcolumn}%
\usepackage{bm}%
\usepackage[utf8]{inputenc}
\usepackage{fdsymbol}

\def\gal{{\tt galacticus}~}
\def\galns{{\tt galacticus}}

\def\cat{Caterpillar }

\def\symphony{Symphony }

\def\col{{\tt colossus}~}

\def\astropy{{\tt astropy}~}

\def\fssigsub{{$f_{\mathrm{s}} \cdot \Sigma_{\mathrm{sub}}$}}
\def\hanetal{Han et al.~}

\newcommand{\reffig}[1]{Figure \ref{#1}}

\newcommand{\refsec}[1]{Section \ref{#1}}
\newcommand{\refappendix}[1]{Appendix \ref{#1}}
\newcommand{\reftable}[1]{Table \ref{#1}}
\newcommand{\refeq}[1]{Eq. \ref{#1}}

\begin{document}

\preprint{APS/123-QED}

\title{Dark Matter Substructure: A Lensing Perspective}

\author{Charles Gannon}
 \altaffiliation{University of California, Merced, 5200 N Lake Road, Merced, CA, 95341, USA}
 \email{cgannon@ucmerced.edu}
\author{Anna Nierenberg}
 \affiliation{University of California, Merced, 5200 N Lake Road, Merced, CA, 95341, USA}
\author{Andrew Benson}
 \affiliation{Carnegie Observatories, 813 Santa Barbara Street, Pasadena, CA 91101, USA}
\author{Ryan Keeley}
 \affiliation{University of California, Merced, 5200 N Lake Road, Merced, CA, 95341, USA}
\author{Xiaolong Du}
 \affiliation{Department of Physics and Astronomy, University of California, Los Angeles, CA 90095, USA}
 \affiliation{Carnegie Observatories, 813 Santa Barbara Street, Pasadena, CA 91101, USA}
\author{Daniel Gilman}
 \affiliation{Department of Astronomy \& Astrophysics, University of Chicago, Chicago, IL 60637 USA}

\date{\today}%

\begin{abstract}
The study of dark matter substructure through strong gravitational lensing has shown enormous promise in probing the properties of dark matter on sub-galactic scales.
This approach has already been used to place strong constraints on a wide range of dark matter models including self-interacting dark matter, fuzzy dark matter and warm dark matter.
A major source of degeneracy exists between suppression of low mass halos due to novel dark matter physics and the strength of tidal stripping experienced by subhalos.
We study theoretical predictions for the statistical properties of subhalos in strong gravitational lenses using the semi-analytic galaxy formation toolkit: \galns.
We present a large suite of dark matter only \gal models, spanning nearly two orders of magnitude in host halo mass (from Milky Way to group mass halos between redshifts from $0.2$ to $0.8$).
Additionally, we include a smaller set of \gal runs with the potential of a central massive elliptical to complement our dark matter only suite of models.
We place particular focus on quantities relevant to strong gravitational lensing; namely the projected number density of substructure near the Einstein radius as function of host stellar mass and redshift.
In the innermost region in projection, we find that our \gal models agrees with N-body simulations within a factor of $\sim 2$ within the Einstein radius.
We find that the addition of a central galaxy suppresses the projected number density of subhalos within in the Einstein radius by around $15\%$ relative to dark matter only simulations.

\end{abstract}

\maketitle

\section{Introduction}\label{sec:introduction}
Cold dark matter (CDM) provides an excellent explanation of the matter distribution on the largest scales, for example the cosmic microwave background \citep{2020} and luminous galaxies \citep[e.g.][]{white1978core,white1991galaxy,de2008high,weinberg2015cold}.
Recently, probes of dark matter have begun pushing the frontier to subgalactic scales without the need for luminous tracers.
In this regime, stellar streams \citep[e.g.][]{bonaca2019spur,10.1093/mnras/stab210,banik2021novel,bovy2017linear,banik2018probing} promise to provide constraints within the local group.
Outside the Local Group, strong gravitational lensing probes the properties of low mass halos~\citep[e.g.][]{vegetti2018constraining,minor2021unexpected,powell2023lensed, mao1998evidence,keeley2024jwst,dike2023strong,gilman2019probing, 10.1093/mnras/stz3480, gilman2022primordial, gilman2023constraining}.%

Halos can be considered in two categories; isolated (field) halos and satellites within `the lensing galaxy' (subhalos) with both categories contributing to the lensing observables.
While the properties of field halos can generally be robustly predicted for a given dark matter model, subhalos, which are gravitationally bound to the lens halo, undergo complex interactions including tidal heating, tidal stripping and dynamical friction, which all occur within the evolving potential of the host halo.
These effects can be degenerate with the effects of dark matter particle physics which can also act to suppress the halo mass function. 
The subhalo to field halo ratio is strongly dependent on the lens and source redshift, with subhalos accounting for $5$--$20\%$ of the total number of halos near the lensed images \citep{10.1093/mnras/stu2673, 10.1093/mnras/sty159}. 
Although subhalos are subdominant in number, uncertainty in the normalization of the subhalo mass function (SHMF) is a major source of degeneracy in strong gravitational lensing studies.  
For example, Gilman et al. \citep{gilman2024turbocharging} simulated 25 mock lenses with warm dark matter (WDM) with a half mode mass ($m_{\mathrm{hm}} = 10^{7.5} \mathrm{M}_\odot$) as a ground truth.
When a Gaussian prior with width 0.2 dex was used, the lower bound on the half-mode-mass improved by 2 dex relative to when a uniform prior with width of 1.5 dex was used.
The latter prior is comparable to what is used in current lensing studies.

This work focuses on studying the statistical properties of low-mass subhalos with a focus on observables related to quadruply lensed quasars, with lenses on the group scale.
Current lensing studies use a wide prior (a factor of $30$) on the normalization of the subhalo mass function of subhalos within the Einstein radius to account for theoretical uncertainties in the tidal evolution of subhalos.
One way to break degeneracies and improve constraining power is to use a narrower prior  based directly on predictions from N-Body simulations.
\citep{gilman2024turbocharging}.
However, caution must be used: a large body of works has drawn attention to artificial disruption present in N-body simulations \citep{10.1093/mnras/stx2956,10.1093/mnras/stz2767,10.1093/mnras/stz3349,2022MNRAS.517.1398B}.
Even in high resolution simulations, such as the \cat \citep{griffen2016caterpillar} suite of N-body simulations of Milky Way mass halos with a per particle mass of $\sim 10^{4} \mathrm{M}_\odot$ artificial disruption can lead to a $10 - 20 \%$ suppression of the SHMF over the entire viral volume \citep{2022MNRAS.517.1398B}.
Additionally, Benson \& Du \citep{2022MNRAS.517.1398B} found significant spatial dependence in the effects of artificial disruption, with the SHMF in the inner $2 \%$ of the virial radius being spuriously suppressed by nearly a factor of $3$.
Due to the geometry of projection, subhalos at small radial displacement from the host are over-represented in the populations of subhalos probed by lensing, making the effects of artificial disruption more pronounced when compared to the total subhalo population.

Due to their computational cost, zoom-in simulations are not conducive to studying population level statistics of subhalos over the large range of parameter space of host halo masses and redshift representative of lenses used to study dark matter.
Analytic and semi-analytic models of subhalo interactions provide a path forward.
These methods must account for the complex interactions between the subhalos and their host. %
Developing fast and accurate models of the impact of environment on subhalos remain an area of active study.
In particular, models of tidal stripping have recently received much attention, which much recent work focused on understanding the tidal tracks of subhalo evolution \citep{2020MNRAS.491.4591E, errani2021asymptotic,2022MNRAS.517.1398B,2024arXiv240309597D}.

In light of the improved understanding of artificial disruption in N-Body simulations, this study re-visits the predictions for the joint spatial and mass distribution of subhalos in CDM, with emphasis on quantities related to gravitational lensing.
We use semi-analytic models as our primary mode of emulating dark matter subhalos.
In particular, we use the \galns
\footnote{https://github.com/galacticusorg/galacticus , we use revisions 723170fb1690257be9b0588c0a83c9e559a584ae and 0e3a5adea6e3b32b9eaaa3a6c4896bbe63fb0cb1 }
\citep{BENSON2012175} galaxy formation toolkit.
The \gal galaxy formation toolkit is a modular and open source semi-analytic model of galaxy formation with an extensive library of both dark matter and baryonic physics.%

We present a large suite of \gal subhalo realizations, spanning nearly two decades in host halo mass (from Milky Way mass to group mass halos at redshifts $0.2$--$0.8$).
We use our suite of \gal runs to study quantities relevant to strong gravitational, with particular focus placed on the subhalo population within a $20$ kpc aperture.
A $20 ~\text{kpc}$ aperture is chosen to be small enough to be comparable to the Einstein radius of a typical SLACS Shu et al. \citep{shu2017sloan} ($\sim 1 ~\text{arcsec}$ corresponding to $\sim 8 ~\text{kpc}$ for a lens at redshift $z = 0.5$), but large enough to obtain reasonable statistics from our \gal models.
To complement our \gal models, we include comparison with the \symphony suite of N-body simulations as well as analytic models such as those given by \hanetal \citep{10.1093/mnras/stv2900}.
Similar to previous works focusing on the subhalo populations of strong gravitational lenses, our main suite of models is limited to dark matter only physics.
We make a preliminary study of baryonic effects by studying a smaller suite of models that include the potential of a central galaxy.
We take an empirical approach to modeling the evolution of the central galaxy using the {\tt UniverseMachine } \citep{Behroozi_2019} correlation between galaxy growth and dark matter halo assembly.

In \refsec{sec_methods}, we discuss the parameters chosen for our \gal models and the methods used in our analysis.
Next, in \refsec{sec_results} we present our results and provide discussion in \refsec{sec_discussion}.
Finally, we summarize our results in \refsec{sec_conclusion}.
For this work, we assume cosmological parameters from the Planck Collaboration \citep{2020}, $(H_0, \Omega_m, \Omega_\Lambda) = (67.36, 0.31530, 0.68470)$.
A combination of \astropy \citep{astropy:2013,astropy:2018,astropy:2022}, \col \citep{diemer2018colossus} and \gal \citep{BENSON2012175} software packages are used for cosmological calculations.
To calculate halo concentrations the Diemer et al. \citep{diemer2019accurate} model is used.
All quantities are reported in physical units unless otherwise specified.

\section{Methods}\label{sec_methods}
In this section we discuss the methods used in this work.
We discuss our semi-analytic models in \refsec{sec_methods_models}.
In \refsec{sec_methods_analytic} we provide an overview of theoretical expectations for the spatial distribution and mass function of subhalos.

\subsection{Semi-Analytic Models}\label{sec_methods_models}
We utilize semi-analytic models as our primary mode of studying the population level statistics of dark matter substructure.
Semi-analytic models have the advantage of speed when compared to N-body simulations, and are chosen to be the primary focus of this work to enable fast exploration of the population level statistics of the substructure of dark matter halos.
In particular, we use \gal, an open source and extensible galaxy formation framework.
The \gal toolkit is particularly well suited to the study of subhalo properties due to extensive library of subhalo physics and its recent calibration to high resolution idealized simulations provided in Du et al. \citep{2024arXiv240309597D}.

Here, we give a brief give a summary of the  algorithms used to generate realizations of dark matter substructure.
We use \gal to generate merger trees and then subsequently evolve the trees forward in time using a set of analytic models for the evolution of subhalo density profiles and orbits (orbit initialization, dynamical friction, tidal stripping and tidal heating).
In \gal, dark matter merger trees are built using Monte Carlo algorithms presented in Cole et al. \citep{10.1046/j.1365-8711.2000.03879.x} and Parkinson et al. \cite{parkinson2008generating}.
In this algorithm, merger trees are evolved backwards in time, with branching rates calculated at each timestep.
Branching rates are calculated at each timestep using extensions \citep{bond1991excursion, bower1991evolution}
to Press \& Schechter \citep{press1974formation} formalism.
Merger tree nodes are then evolved forward in time.
First, orbits are initialized using the potential of the host halo \citep{jiang2015orbital}.
After initialization, subhalo orbits are then evolved using models for dynamical friction \citep{chandrasekhar1943dynamical}, tidal stripping \citep{zentner2005physics} and tidal heating \citep{gnedin1999tidal}.

We use subhalo physics tuned to high resolution idealized simulations provided by Du et al. \citep{2024arXiv240309597D} in which subhalos are evolved in an analytic host potential.
When compared to tradition cosmological zoom-in simulations, these simulations have the advantage of resolving subhalos at a much higher resolution for less computational cost.
For example, in the idealized simulations used by Du et al. \citep{2024arXiv240309597D}, subhalos start with $N \approx 10^7$ particles and are tracked until $N \approx 10^4$ particles remain.
This is in contrast to high resolution cosmological such as \cat \citep{griffen2016caterpillar}, where the lowest mass subhalos are tracked to $N \approx 20$ particles.
The high resolution enabled by using idealized simulations minimizes the effects of numerical issues traditionally associated with N-body simulations, such as artificial disruption \citep{10.1093/mnras/sty084}.
To further account for the effects of artificial disruption, even when tuning to cosmological simulations, a future calibration of \gal is planned using the methods presented in Benson \& Du \citep{2022MNRAS.517.1398B}.
Additionally, the challenges associated with halo finding in cosmological simulations is not present in idealized simulations.
However, we note that our idealized simulations have several limitations; idealized simulations lack cosmological context and use simplified, spherically symmetric analytic profiles to describe the host halo and infalling subhalo.

Probes of strong gravitational lensing are sensitive to subhalos appearing in projection near lensed images which appear near the Einstein radius.
For a typical group mass lens at redshift $z=0.5$, this aperture has a radius of $\sim 1$ arcsecond ($\sim 7 ~\text{kpc}$ \citep[][]{shu2017sloan})
Due to effects of projection, strong lensing probes are especially sensitive to subhalos at small spatial separation to the host.
The \gal galaxy formation model automatically destroys subhalos within a minimum radius.
We select the minimum radius to simulate subhalos by running convergence tests.
We find convergence for the spatial distribution at $10$ kpc for a merging radius of $0.01~r_{\mathrm{v}}$.
For a $10^{13.5} \mathrm{M}_{\odot}$ mass halo at redshift, z = 0.2, $0.01~r_{\mathrm{v}}$ corresponds to a radial separation from the center of the host of $0.7$ kpc.
Unless otherwise noted, when considering projected quantities, we include subhalos within an annulus with an inner and outer radius of $10$ and $20$ kpc respectively.
This ensures that the subhalo population within $10$ kpc, which may be incomplete, is excluded.
We run our models with a tree (infall mass) floor of $8 \times 10^7 \
\mathrm{M}_\odot$ to ensure completeness of the population of subhalos with infall mass $m > 10^{8} \mathrm{M}_{\odot}$.
We track subhalos to arbitrarily low bound mass, ensuring no subhalos are destroyed by tidal stripping.
We model across a broad range of masses and redshifts to capture the evolution of the projected number density and to facilitate comparison with observational works as well as other simulations.
Our suite of \gal models consists of a 10 by 10 grid of \gal outputs spanning the range of host halo masses $12.0 \leq \log_{10}\left( M_{\mathrm{h}}/ \mathrm{M}_\odot \right ) \leq 13.5$ and redshifts $0.2 \leq z \leq 0.8$ with 224 host halos per grid point.
In addition to our dark matter only models, we include a smaller suite of \gal models spanning $13.0 < \log_{10} (M_{\mathrm{h}} / \mathrm{M}_\odot) < 13.5$ with the inclusion of the potential of a central massive elliptical.
Along with the increased tidal heating and stripping due to the potential of the central galaxy, we include a model of baryonic contraction due to the central potential \citep{gnedin2004response}.
The morphology of the central massive elliptical is modeled using a Hernquist \citep{hernquist1990analytical} profile.

We take care to accurately model the evolution of the stellar mass and scale radius of the central galaxy.
We model the evolution of the central galaxy using the stellar mass to halo mass relation of Behroozi et al. \citep{Behroozi_2019}, Equation J3.
For our work, we use the best fit parameters provided in the first row of table J1.
For simplicity, we do not include intrinsic scatter in the stellar mass to halo mass relation.
We use empirical power law fits provided by Shen et al. \citep{2003MNRAS.343..978S} for the stellar mass to stellar radius relationship for early type galaxies in the Sloan Digital Sky Survey \citep{york2000sloan} to evolve the Hernquist radius of the central galaxy.
Similarly, we do not include scatter in this relationship.
We provide further discussion of the evolution of the central galaxy in \refappendix{sec:appendix}.

\subsection{Analytic Models}\label{sec_methods_analytic}
Here, we discuss theoretical expectations for the subhalo population, with emphasis placed on the spatial and mass distribution of subhalos.
Two cases of the spatial-mass distribution can be considered, using two definitions of subhalo mass: the mass at infall and the gravitationally bound mass of the subhalo.
The former is known as the ``unevolved'' distribution and the latter the ``evolved'' distribution.
By definition, no mass loss due to tidal stripping is considered in the unevolved distribution, while the evolved distribution includes tidal mass loss.
Due to the extreme mass loss undergone by some subhalos, the unevolved distribution is not an observable quantity.
Instead, the unevolved distribution provides a powerful tool to study the properties of subhalos, allowing the complicated physics of tidal stripping to be separated from orbital physics.
A useful picture is that the evolved distribution can be approximately thought of as being derived from the unevolved distribution and a model of tidal mass loss as shared in \hanetal \citep{10.1093/mnras/stv2900} and Gilman et al. (In prep).
However, we note that this picture is not exact. 
For example, physics of tidal stripping cannot be completely separated from orbital physics as tidal mass loss can affect dynamical friction (\gal uses the bound mass in dynamical friction calculations).

To analytically describe the spatial and mass distribution, we use the  unified model presented in \hanetal \citep{10.1093/mnras/stv2900}.
We compare the predictions from N-body simulations and semi-analytic models with the analytic prescription presented by \hanetal \citep{10.1093/mnras/stv2900} which assumes that the unevolved spatial-mass distribution of subhalos is separable (the spatial and mass function can be separated into a mass function and spatial distribution with the mass function only dependent on the halo mass and spatial distribution only dependent on the subhalo's radial separation from the host), and that the unevolved spatial distribution of subhalos traces the density profile of the host.
Under the assumption of separability, the unevolved spatial and mass function can be written as
\begin{equation}\label{eq_han_unevolved}
    \frac{\mathrm{d}^{2}N}{\mathrm{d}V\mathrm{d}m} = \Tilde{\rho}(r) \frac{\mathrm{d}N}{\mathrm{d}m}(m),
\end{equation}
where $\frac{\mathrm{d}^{2}N}{\mathrm{d}V\mathrm{d}m}$ is the unevolved subhalo spatial and mass function, $\tilde{\rho}$ is the normalized density profile of the host and $\mathrm{d}N/\mathrm{d}m$ is the unevolved mass function.
\hanetal \citep{10.1093/mnras/stv2900} then model the evolved spatial and mass distribution as separable, with the unevolved spatial distribution rescaled by a radially dependent transfer function $\hat{T}(r)$
\begin{equation}\label{eq_han_evolved}
    \frac{\mathrm{d}^{2}N}{\mathrm{d}V\mathrm{d}m_{\mathrm{b}}} = \hat{T}(r) \Tilde{\rho}(r) \frac{\mathrm{d}N}{\mathrm{d}m_{\mathrm{b}}},
\end{equation}
where $\frac{\mathrm{d}^{2}N}{\mathrm{d}V\mathrm{d}m_{\mathrm{b}}}$ is the evolved spatial and mass function, $\mathrm{d}N/\mathrm{d}m_{\mathrm{b}}$ is the evolved mass function and $m_{\mathrm{b}}$ is the gravitationally bound mass of the subhalo.
In their analysis, \hanetal \citep{10.1093/mnras/stv2900} take $\hat{T}(r)$ to be a power law
\begin{equation}\label{eq_han_transfer}
    \hat{T}(r) = \left(\frac{r}{r_{\mathrm{v}}}\right)^{\gamma},
\end{equation}
where $\gamma \approx 1$ is a free parameter.
\hanetal \citep{10.1093/mnras/stv2900} fit $\gamma$ to various N-body simulations finding $\gamma = 0.95$ for Milky way mass Aquarius halos \citep{10.1111/j.1365-2966.2008.14066.x} and $\gamma = 1.33$ for group mass halos in the Phoenix suite \citep{2016ApJ...824..144F}, indicating a dependence on host halo mass.
The unevolved SHMF is taken to be a power law,
\begin{equation}\label{eq_han_massfunction_unevolved}
    \frac{\mathrm{d}N}{\mathrm{d}m} \propto m^\alpha,
\end{equation}
where $\alpha \approx -2$ is the logarithmic slope of the SHMF.
Directly following from the assumption of separability, the evolved SHMF is predicted to follow a power law with identical slope.

To analyze our \gal results, we fit the \hanetal \citep{10.1093/mnras/stv2900} models to the \gal predictions for the spatial and mass function in \refsec{sec_results}.
To provide an additional point of comparison to our \gal results in the context of gravitational lensing, we spatially project the \hanetal \citep{10.1093/mnras/stv2900} into 2d.
An immediate consequence of the separability assumed by the  \hanetal \citep{10.1093/mnras/stv2900} model is a prediction that the logarithmic slope of the SHMF should be identical between the projected and unprojected distributions.
Additionally, we study the predictions of the model for the scaling of the projected subhalo mass function (PSHMF) within the Einstein radius, an important quantity for gravitational lensing.
We estimate the scaling of the projected spatial and mass function as a function of host halo mass utilizing the host halo mass scaling relations discussed in \hanetal \citep{10.1093/mnras/stv2900}.
However, redshift dependent scaling is not considered by \hanetal \citep{10.1093/mnras/stv2900}, who derived their model for a single snapshot in time at $z = 0$.
Typical lensing galaxies are found at redshifts $0.2 \lesssim z \lesssim 0.8$, so understanding the scaling of the SHMF with redshift is essential.
Therefore, we discuss extending the \hanetal \citep{10.1093/mnras/stv2900} model over these range of redshifts in the following sections.

\section{Results}\label{sec_results}

Here, we analyze our suite of \gal models.
In \refsec{sec_radial} we discuss the spatial distribution predicted by \galns.
Next, in \refsec{sec_norm} we calculate the normalization of the PSHMF of our dark matter only (DMO) simulations, and repeat the calculations for our models with the potential of a central massive elliptical in \refsec{sec_galaxy_norm}.
Next, in \refsec{sec_scaling} we predict the scaling of the PSHMF as a function of host halo mass and redshift using our dark matter simulations. 
Finally, we discuss the impact of a central galactic potential on the scaling of the PSHMF.

\subsection{Radial Distribution}\label{sec_radial}

\begin{figure*}
    \includegraphics[width=\textwidth]{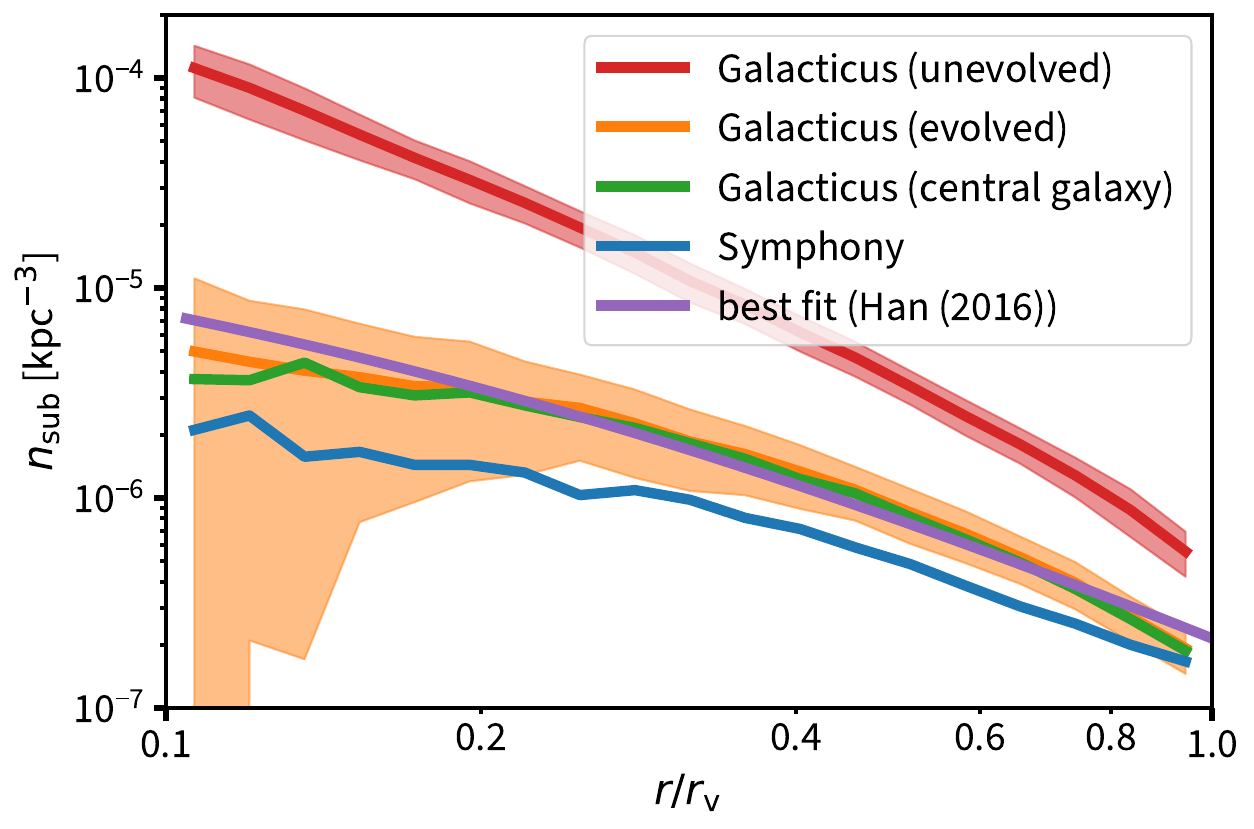}
    \caption{
                The differential spatial distributions ($n_{\mathrm{sub}}$) from our \gal models and Symphony for subhalos with mass $9 < \log_{10}\left (m_{b}/\mathrm{M}_{\odot} \right ) < 10$ plotted as a function of the virial radius fraction ($r / r_{\mathrm{v}}$).
                Results are for halos with mass $M_{\mathrm{h}} = 10^{13} \mathrm{M}_\odot$ at redshift $z=0.5$.
                In red: the unevolved spatial distribution from our \gal models, in orange: the evolved spatial distribution for \gal and in green with a central galaxy.
                The shaded regions show the $1 \sigma$ halo to halo scatter for the \gal evolved and unevolved case, scatter for the other distributions is not shown for visual clarity. 
                The best fit of the Han (2016) model to the \gal DMO unevolved spatial distribution is shown in purple.
                In blue: the spatial distribution for Symphony.
            }
    \label{fig_spatial_3d}
\end{figure*}

\begin{figure*}
  \includegraphics[width=\textwidth]{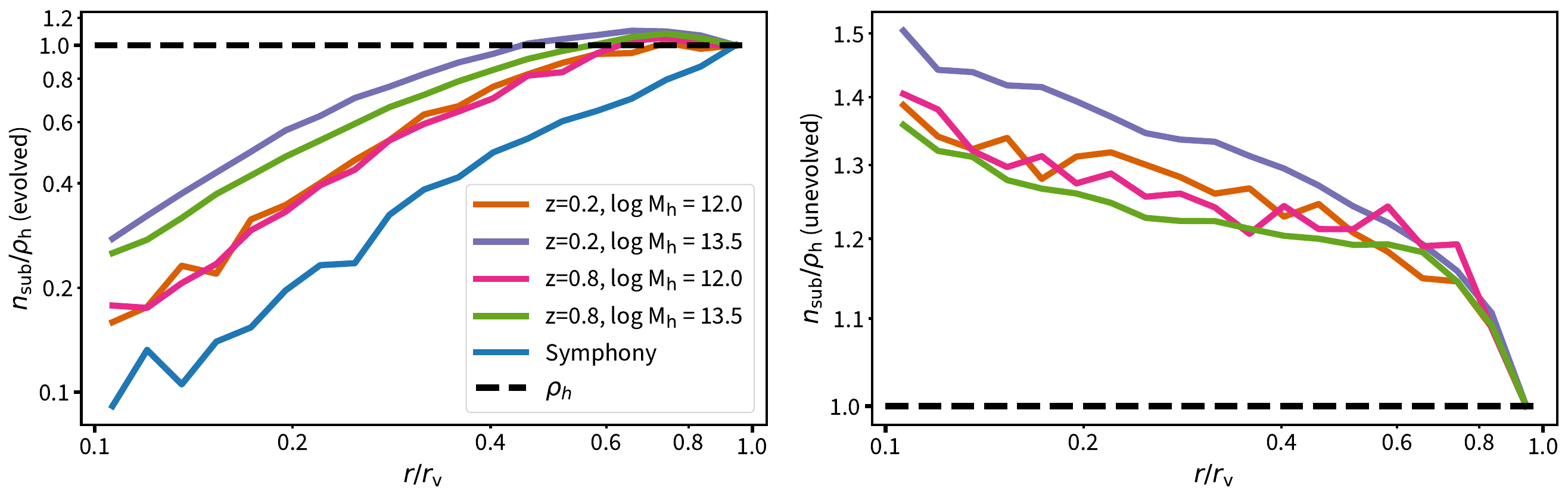}
  \caption{%
        The ratio of the \gal spatial distribution ($n_{\mathrm{sub}}$) to the host's density profile ($\rho_{\mathrm{h}}$) for subhalos with $8 < \log_{10}(m_{b} / \mathrm{M}_\odot) < 9$ normalized to $1$ at $r = r_{\mathrm{v}}$ for the evolved distribution (left) and $8 < \log_{10}(m / \mathrm{M}_\odot) < 9$ unevolved distribution (right).
        A horizontal line of 1 (black dotted) shows the host halos density profile.
        Results are plotted for \gal over a range of host halo masses and redshifts and for Symphony for a single mass and redshift ($\log_{10} \left ( M_{\mathrm{h}} / \mathrm{M}_\odot \right ) = 13, z=0.5$).
        Symphony results are plotted for subhalos in the mass range $8 < \log_{10}(m_{b} / \mathrm{M}_\odot) < 9$.
        Over our range of redshifts explored by our \gal simulations, the slope of the evolved spatial distribution $n_{\mathrm{sub}} / \rho_{\mathrm{h}}$ has a mild dependence on halo mass, with almost no dependence on redshift.
        For the unevolved case, there appears to be no significant dependence one either redshift or host halo mass.
   }
  \label{fig_spatial_ratio}
\end{figure*}

\reffig{fig_spatial_3d} shows the spatial distribution predicted by \gal for a $10^{13} \mathrm{M}_\odot$ halo at $z=0.5$, with both the evolved and unevolved distributions plotted.
Alongside the \gal prediction, results are shown for \symphony \citep{nadler2023symphony} and \hanetal \citep{10.1093/mnras/stv2900} model of the spatial distribution.
We plot the ratio of the radial distribution of subhalos to the host's dark matter density in \reffig{fig_spatial_ratio}.
For a $10^{13} \mathrm{M}_\odot$ halo at $z=0.5$ we find the best fit to the \gal spatial distribution with a value of $\gamma = 0.98$.
We fit the spatial distributions predicted by \gal with \hanetal \citep{10.1093/mnras/stv2900} model, and summarize the best fit $\gamma$ values in \reftable{tab_han_fits}. 
Additionally, fits provided by \hanetal \citep{10.1093/mnras/stv2900} to Aquarius \citep{10.1111/j.1365-2966.2008.14066.x} and Phoenix~\citep{2016ApJ...824..144F} are included for comparison in \reftable{tab_han_fits}.

The fits to the $10^{12} \mathrm{M}_\odot$ \gal halos agree well with fits to Aquarius halos ($\approx 10^{12} \mathrm{M}_\odot$).
Additionally, we find a trend of increasing $\gamma$ with host halo mass in our suite of \gal models, similar to that found by \hanetal \citep{10.1093/mnras/stv2900} when comparing the Aquarius and Phoenix simulations ($\approx 7 \times 10^{14} \mathrm{M}_\odot$).
Over the range of redshifts probed $0.2 \leq z \leq 0.8$, and at a fixed halo mass we find that $\gamma$ is nearly constant.
Further discussion of the \hanetal \citep{10.1093/mnras/stv2900} model is provided in \refsec{sec_discuss_pop}.

\begin{table*}
    \centering
    \begin{tabular}{ccccc}
        \hline
        Simulation Suite / Model    &  Source                       & $\log_{10}(M_{\mathrm{h}} / \mathrm{M}_\odot)$    & $z$   & $\gamma$  \\
        \hline
        \gal                        & this work                     & $12.0$                        & 0.2   & 0.94      \\
        \gal                        & this work                     & $12.0$                        & 0.8   & 0.90      \\
        \gal                        & this work                     & $13.0$                        & 0.5   & 0.98      \\
        \gal                        & this work                     & $13.5$                        & 0.2   & 1.23      \\
        \gal                        & this work                     & $13.5$                        & 0.8   & 1.24      \\
        Aquarius                    & \hanetal \citep{10.1093/mnras/stv2900}  & $12.0$                        & 0.0   & 0.95      \\
        Phoenix                     & \hanetal \citep{10.1093/mnras/stv2900}  & $14.8$                        & 0.0   & 1.33      \\
        \hline
    \end{tabular}
    \caption{
        Table of best fits to \hanetal model of the spatial distribution (see \refeq{eq_han_evolved}).
        We include fits spanning the range of halo masses and redshifts included in our suite of \gal models, as well as at  the redshift and host halo mass, $M_h = 10^{13} \mathrm{M}_\odot$ and redshift, $z = 0.5$ where $\Sigma_{\mathrm{sub}}$ is defined.
        For comparison, we include fits to Aquarius and Phoenix halos provided by \hanetal.
    }
    \label{tab_han_fits}
\end{table*}

\begin{figure*}
    \includegraphics[width=\linewidth]{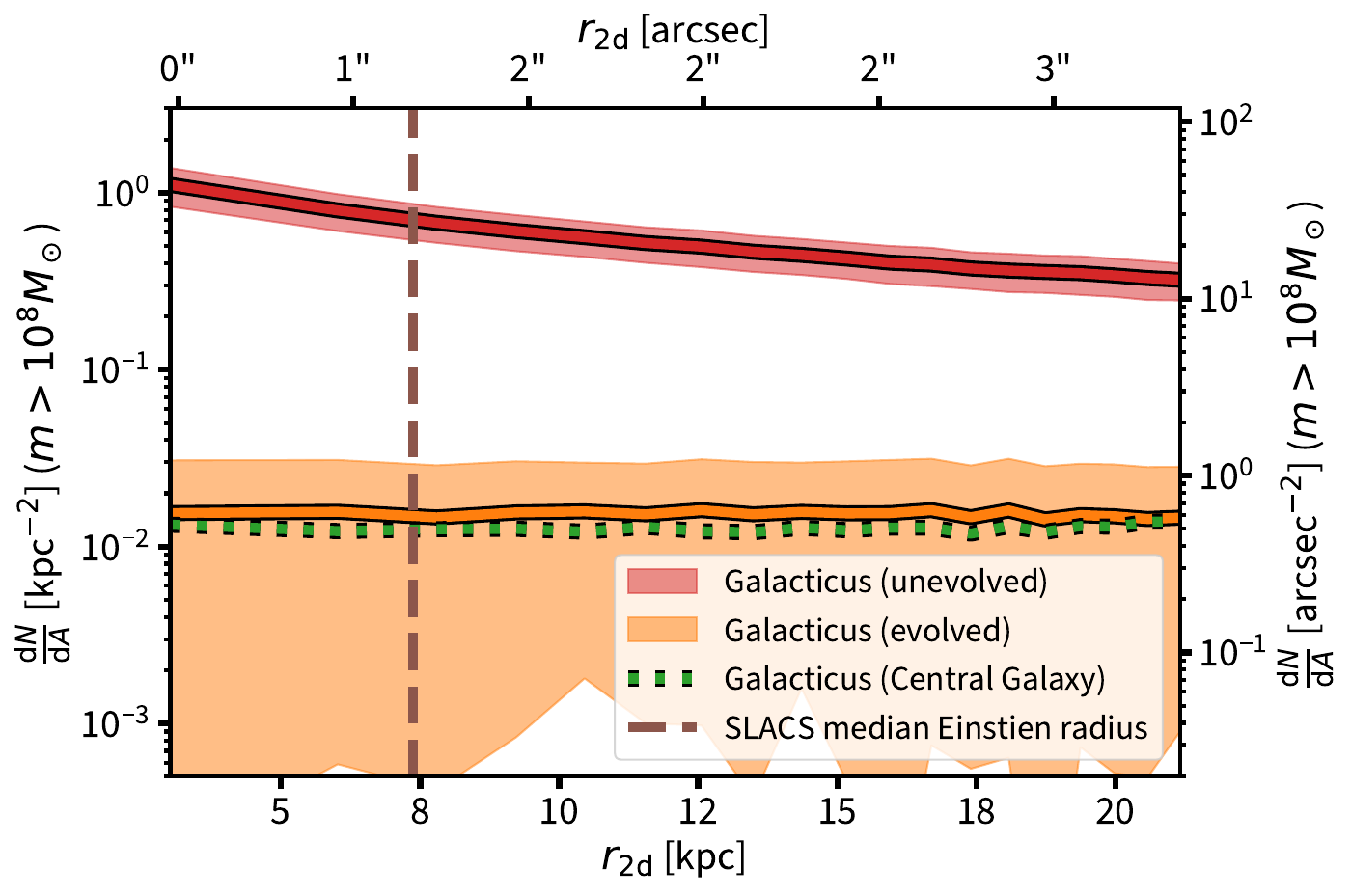}
    \caption{
                The projected number density within the inner 20 kpc for both the unevolved (upper red), evolved with central galaxy (green dotted) and evolved (lower orange) subhalo distributions.
                Predictions from \gal are shown for subhalos with mass $> 10^{8} \mathrm{M}_\odot$ in a $10^{13} \mathrm{M}_\odot$ halo, with scatter shown for 76 kpc$^2$ bins.
                In brown (dotted) the Einstein radius for a typical SLACS lens is shown.
                The evolved number density is constant within the innermost region of the halo for both DMO and central galaxy models.
                The shaded region shows $1 \sigma$ halo to halo scatter, with scatter not being shown for the case with a central galaxy for visual clarity.
    }
    \label{fig_spatial_2d_multihalo}
\end{figure*}

The projected spatial distribution of subhalos in the mass range $10^8 \mathrm{M}_\odot < m < 10^9 \mathrm{M}_\odot$ predicted by \gal for a $10^{13} \mathrm{M}_\odot$ halo at $z=0.5$ with projected radii $r_{\mathrm{2d}} < 20 ~\text{kpc}$ is shown in
\reffig{fig_spatial_2d_multihalo}.
For reference, we show the median Einstein radius from the sample of SLACS lenses presented by Shu et al. \citep{shu2017sloan}.
We find that the projected spatial distribution is nearly constant in the innermost region of the host, which is consistent with the findings of Xu et al. \citep{10.1093/mnras/stu2673}.
We find that this holds true for both our DMO models and models with a central galaxy.

\subsection{Projected Subhalo Mass Function Normalization---Dark Matter Only}\label{sec_norm}
Of particular interest to lensing studies is the integrated number of subhalos within an aperture near the Einstein radius as a function of the properties of the host halo.
To measure this, we use the convention of Gilman et al. \citep{10.1093/mnras/stz3480} to define the average density of the unevolved projected subhalo mass function (PSHMF) within the Einstein radius as:
\begin{equation}\label{eq_massfunction_proj}
    \frac{\mathrm{d}^2N}{\mathrm{d}m\mathrm{d}A} =  \frac{\Sigma_{\mathrm{sub}}}{m_0}\left (\frac{m}{m_0} \right)^{\alpha}F(M_{halo},z),
\end{equation}
where $\frac{d^2N}{dmdA}$ is a measure of subhalo number density, $m$ is the mass of a subhalo at infall, $m_{0}$ is a pivot mass, taken by convention to be $10^{8} \mathrm{M}_{\odot}$, and the function $F(M_{halo},z)$ describes the dependence on host halo mass and redshift.
For convenience, we will refer to $\frac{d^2N}{dmdA}$ as the projected subhalo mass function (PSHMF).
Factoring out all dependence on redshift and host halo mass into the scaling function $F(M_{halo},z)$ allows a single PSHMF normalization ($\Sigma_{\mathrm{sub}}$) to be measured for all host halos regardless of host halo mass or redshift.
To provide further context for \refeq{eq_massfunction_proj}, we can define the evolved PSHMF:
\begin{equation}\label{eq_massfunction_proj_evolved}
    \frac{\mathrm{d}^2N}{\mathrm{d}m_{\mathrm{b}}\mathrm{d}A} = \frac{f_{s} \cdot \Sigma_{\mathrm{sub}}}{m_0} \left( \frac{m_{b}}{m_0} \right)^{\alpha_{\mathrm{b}}} F_b(M_{halo},z),
\end{equation}
where $f_{s}$ accounts for the reduction in the normalization of the SHMF from tidal stripping, $\alpha_{\mathrm{b}}$ is the logarithmic slope of the evolved SHMF and $m_{b}$ is  the bound mass of the halo.

Eq. \ref{eq_massfunction_proj} and \ref{eq_massfunction_proj_evolved} can be derived by spatially projecting the \hanetal \citep{10.1093/mnras/stv2900} models of spatial and mass distribution into 2d.
An immediate consequence of the mass independent model of tidal stripping assumed by \hanetal \citep{10.1093/mnras/stv2900} is that the logarithmic slopes of the evolved and unevolved mass functions should be identical.
Additionally, under the separability of the spatial and mass function assumed by \hanetal \citep{10.1093/mnras/stv2900}, the logarithmic slopes of the total and projected mass functions should be identical.
We check these assumptions against our model results in \reftable{tab_scaling} and discuss why this assumption may be inaccurate due to mass segregation due to dynamical friction in the unevolved distribution in section \refsec{sec_discuss_pop}.
However, for simplicity, in the remainder of this work we assume $\alpha = \alpha_{\mathrm{b}} = -1.93$ in accordance with the evolved results tabulated in \reftable{tab_scaling}.

All dependence on halo mass and redshift in \refeq{eq_massfunction_proj} and \refeq{eq_massfunction_proj_evolved} is captured in $F$ and $F_b$ respectively, making $\Sigma_{\mathrm{sub}}$ and $f_{\mathrm{s}}$ independent of halo mass and redshift.
Fits of the \hanetal model to the \gal spatial distribution are given in \reftable{tab_han_fits}.
Best fits for $\Sigma_{\mathrm{sub}}$, $f_{\mathrm{s}}$ and $\alpha$ for subhalos within an annulus with inner and outer radii of $10$ kpc and $20$ kpc are given in \reftable{tab_fits}.
We choose A $10-20$ kpc annulus to exclude the innermost subhalo population which may be subject to destruction due to satellite merging implemented in \galns.

Note that because the evolved projected spatial distribution is nearly independent of radius near the center of the host, the normalization of the evolved mass (density) function (\fssigsub) will be nearly independent of the choice of aperture radius.
We compare \fssigsub~to the PonosV~\citep{2016ApJ...824..144F} and PonosQ~\citep{2016ApJ...824..144F} simulations in \reftable{tab_fits_external}. 

\subsection{Projected Subhalo Mass Function Normalization---Impact of Central Galaxy}\label{sec_galaxy_norm}
\begin{figure*}
    \includegraphics[width=\textwidth]{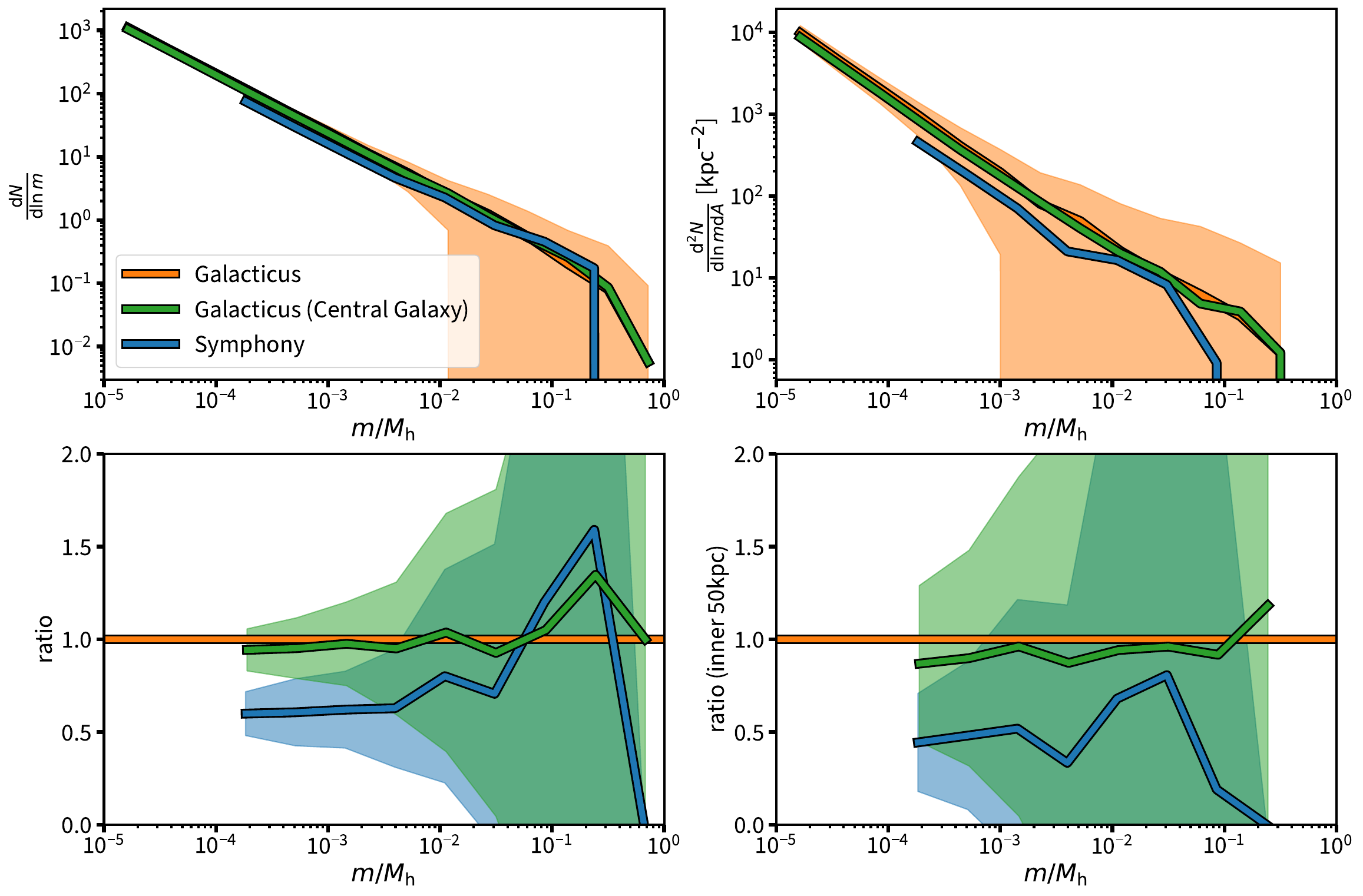}
    \caption{
             Upper left: the SHMF over the entire virial volume, right: the SHMF for subhalos projected within $50$ kpc.
             The \gal DMO predictions are shown in orange, \gal predictions for the central galaxy are shown in green, and results for the symphony suite are shown in blue.
             Lower left and right show the ratios to the \gal DMO predictions.
             The effect of the central galaxy can be seen as a reduction in normalization of the SHMF no significant change to the logarithmic slope, at least for low mass halos.
             Note that the difference in normalization for \gal with and without a central galaxy is significantly less than the difference between our \gal models and the Symphony dark matter only simulation suite. 
     }
    \label{fig_massfunction}
\end{figure*}

\reffig{fig_spatial_3d} shows the impact of the galaxy on the spatial distribution, while \refappendix{sec:appendix} gives more information on the evolution of the central galaxy in our \gal models.
Our \gal models predict the central galaxy has a minimal impact on the SHMF, as shown in \reffig{fig_massfunction}. Within the inner $20$ kpc in projection, both the DMO and galactic potential results have no dependence on radius as shown in \reffig{fig_spatial_2d_multihalo}. For low mass halos, our models predict the central galaxy introduces a mass independent rescaling of the SHMF, with no change in slope. The rescaling of the SHMF is dependent on the distance from the host, with the SHMF in the inner region more heavily suppressed when compared to the DMO predictions.
For a $10^{13} \mathrm{M}_\odot$ halo, the SHMF is suppressed at the $ < 5\%$ level over the entire virial volume when compared to DMO predictions.
In the inner $10-20$ kpc annulus the suppression increases to $15\%$.
We find the impact of the central galaxy is minor when compared to theoretical uncertainties in the SHMF, the difference in normalization between \gal and Symphony is greater than the difference between \gal with and without a central galaxy.

\subsection{Projected Subhalo Mass Function Scaling---Dark Matter Only}\label{sec_scaling}

\begin{figure*}
    \includegraphics[width=\textwidth]{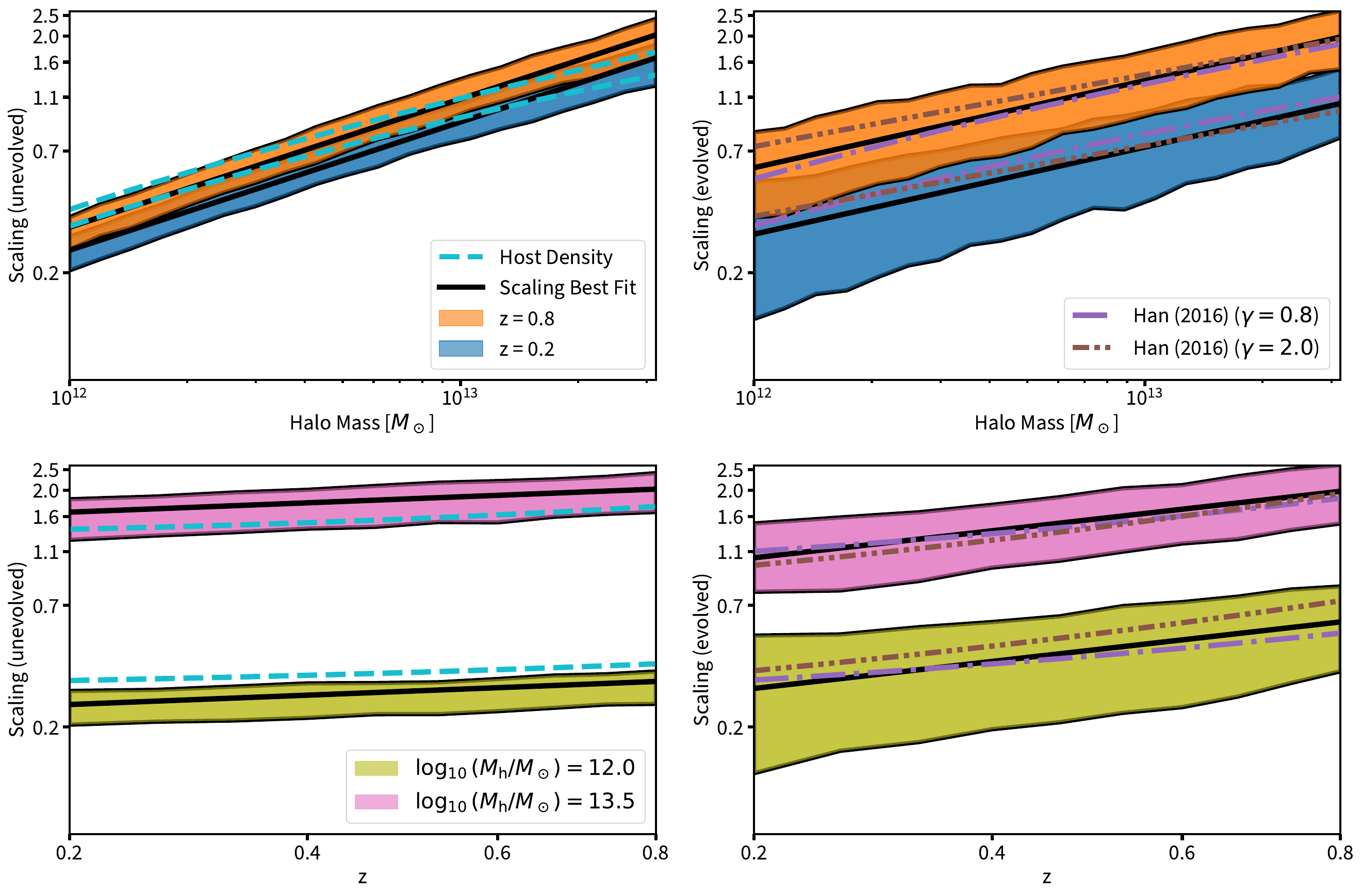}
    \caption{
        Projected number density scaling as a function of host halo mass and redshift for an annulus with an inner and outer radius of $10$ and $20$ kpc respectively.
        The left panels show the unevolved subhalo scaling, while the right panel shows the scaling of the evolved distribution.
        Upper panels show the halo mass scaling relation for host halos at redshift $z=0.2$ (blue) and $z=0.8$ (orange).
        Lower panels show the redshift scaling relation for a host halo mass of $10^{12.0} \mathrm{M}_\odot$ (olive) and $10^{13.5} \mathrm{M}_\odot$ (pink).
        Shaded regions show the $1 \sigma$ halo to halo scatter.
        The unevolved scaling is compared to the scaling of the host's density profile (adjusted to the definition of halo mass used here) in blue (dashed).
        The scaling relations derived from the \hanetal analytic number density distribution (Eq. \ref{eq_han_unevolved} and \ref{eq_han_evolved}) are shown for $\gamma = 0.8$ (purple, dot dashed) and $\gamma = 2.0$ (brown, dot dashed).
        These values of $\gamma$ span a range of physically plausible values.
    }
    \label{fig_scaling}
\end{figure*}

To parameterize $F$ and $F_b$ we follow the procedure used by Gilman et al. \citep{10.1093/mnras/stz3480} to model $F$ and $F_b$ using a power law expansion
\begin{equation}\label{eq_dnda_scaling}
    F(M_{halo},z) = \left(\frac{M_{halo}}{10^{13}\mathrm{M}_{\odot}}\right)^{k_1}\left(z+0.5\right)^{k_2}
\end{equation}
normalized to $1$ at $M_{\mathrm{h}} = 10^{13} \mathrm{M}_\odot$ and $z=0.5$.
We fit $k_1$ and $k_2$ using our suite of \gal models.
We plot the scaling relation in \reffig{fig_scaling}, and give scaling coefficients in \reftable{tab_scaling}.
We find best fit values of $k_1 = 0.55$ (halo mass scaling coefficient) and $k_2 = 0.37$ (redshift scaling coefficient) for the unevolved distribution and $k_1 = 0.37$ and $k_2 = 1.05$ for the evolved distribution.

The scaling of the PSHMF depends on both the spatial and mass distributions.
To model the PSHMF scaling analytically, we project the~\hanetal \citep{10.1093/mnras/stv2900} model. 
To scale the normalization of the total SHMF (over the entire virial volume) as a function of host halo mass and redshift we use the relations provided by Van Den Bosch et al. ~\citep{10.1111/j.1365-2966.2005.08964.x} for the evolved case and Gao et al. ~\citep{10.1111/j.1365-2966.2004.08360.x} for the unevolved case.
We project \refeq{eq_han_unevolved} and \ref{eq_han_evolved} according to the projection equation:
\begin{equation}
    \frac{\mathrm{d}^2N}{\mathrm{d}A \mathrm{d}m}(r_{2d}) = 2 \int_{r_{2d}}^{r_{\mathrm{v}}} \frac{u}{\sqrt{u^2 + r_{2d}^2}} \frac{\mathrm{d}^2 N}{\mathrm{d}V\mathrm{d}m}(u) \mathrm{d}u,
\end{equation}
assuming a spherically symmetric distribution of subhalos, so that the number density of subhalos depends only on the distance from the center of the host, $r$.
Using the formula provided in Van Den Bosch et al. \citep{10.1111/j.1365-2966.2005.08964.x}, the amplitude of the total unevolved SHMF scales as a function of mass and redshift according to the relation:
\begin{equation}\label{eq_scaling_unevolved}
     \frac{\mathrm{d}N}{\mathrm{d}(m / M_{\mathrm{h}})} = C_u \left(\frac{m}{M_{\mathrm{h}}}\right)^{\alpha},
\end{equation}
where $m$ is the mass of the subhalo (at the time of accretion), $M_{\mathrm{h}}$ is the mass of the host halo, $ \alpha $ is the logarithmic slope and $ C_u $ is the normalization of the unevolved SHMF.
For the evolved SHMF, we scale the amplitude using the empirical formula provided by Gao et al. \citep{10.1111/j.1365-2966.2004.08360.x}:
\begin{equation}\label{eq_scaling_evolved}
    \frac{\mathrm{d}N}{\mathrm{d}m_{\mathrm{b}}} = C_\mathrm{e} M_{\mathrm{h}} f(m_{\mathrm{b}},z) m_{\mathrm{b}}^\alpha,
\end{equation}
where $C_\mathrm{e}$ is the normalization of the evolved SHMF, $m_{\mathrm{b}}$ is the bound mass of the subhalo and $f(m_{\mathrm{b}},z)$ abundance of halos with mass $m_{\mathrm{b}}$ at redshift $z$ per unit mass in the universe.
We compute $f(M_{\mathrm{h}}, z)$  using the Sheth \& Tormen \citep{sheth1999large} mass function.

To calculate the projected scaling relation, we numerically integrate \refeq{eq_han_evolved}.
We first fix $\gamma$  as constant for all host halo masses, and calculate the projected scaling relations in halo mass and redshift.
The range $0.8 \leq \gamma \leq 2.0$ is chosen to represent a reasonable range of possible values of $\gamma$, since $\gamma \approx 1$ is expected.
The limit as $\gamma$ goes to 0 gives a NFW profile (which is the same as the unevolved case).
For $\gamma > 3$, the spatial distribution is increasing as a function of r for all values of r, which is unphysical.
In addition to a static value of $\gamma$, we consider an additional case where $\gamma$ varies as a function of halo mass (``$\gamma$ interp'').
In this case, we linearly interpolate $\gamma$ as a function of halo mass, using fits to the spatial distribution predicted by \gal (see \reftable{tab_han_fits}).
We summarize our scaling models by providing fits to  \refeq{eq_dnda_scaling} in \reftable{tab_fits}. We also use these scaling relations in \reftable{tab_fits_external} to compare our measured values to results from the Aquarius \citep{10.1111/j.1365-2966.2008.14066.x} and Phoenix \citep{10.1111/j.1365-2966.2012.21564.x} simulations as well as the Milky Way prior used in Nadler et al.~\citep{nadler2021dark}.

\subsection{Projected Subhalo Mass Function Scaling---Impact of central galaxy}\label{sec_galaxy_scaling}

We find the central galaxy has only a minor impact on the scaling of the amplitude of the PSHMF in redshift and host halo mass (see \reftable{tab_scaling}).
This is expected given the minimal impact to the spatial distribution.
For more massive halos ($M_{h} > 10^{13} \mathrm{M}_\odot$), the difference when compared to the Dark Matter Only (DMO) model decreases, which is expected due to the decreasing baryon fraction of the central galaxy (see \reffig{fig_um_evolution}).
Unless otherwise stated, for the remainder of this work we use the DMO scaling relation. 

\begin{table*}

 \begin{tabular}{ccccc}
  \hline
    Parameters               & Description & Value     & Halo To Halo Scatter      & Units                     \\
  \hline
    $\Sigma_{\mathrm{sub}}$           & Unevolved PSHMF amplitude          & 387       & 71.2                      & $10^{-3}$ kpc$^{-2}$                \\
    $f_{\mathrm{s}}$                    & Tidal stripping mass fraction      & 0.0282    & 0.0105                    & -                         \\
    $f_{\mathrm{s}} \cdot \Sigma_{\mathrm{sub}}$ & Evolved PSHMF amplitude            &10.7      & 3.81                      & $10^{-3}$ kpc$^{-2}$                \\
    $\alpha$                 & Unevolved logarithmic PSHMF slope  &-1.93    & -                         & -                         \\
    $\alpha_{\mathrm{b}}$               & Evolved logarithmic PSHMF slope    &-1.94    & -                         & -                         \\
    \hline
  \end{tabular}
  \caption{
        Table of best fits for the projected subhalo mass function (PSHMF) (Eq. \ref{eq_massfunction_proj} and \ref{eq_massfunction_proj_evolved}).
        Fitting coefficients are averaged over 224 host halos per redshift and host halo mass (22,400 host halos in total).
        Projections are taken within a $10-20$ kpc annulus, and averaged over 3 projections in the xy, yz, and xz planes.
        Due to the low number of halos in the aperture, halo to halo scatter for the SHMF slope cannot be calculated.
 }
 \label{tab_fits}

\end{table*}

\begin{table*}
    \centering
    \begin{tabular}{cccccc}
        \hline
         Simulation & M$_{h}$ [$\mathrm{M}_{\odot}$] & z & PSHMF Amplitude [$10^{-3}\times$~kpc$^{-2}$] & scaled [$10^{-3}\times$~kpc$^{-2}$]& source \\
         \hline
         \gal & $1.0 \cdot 10^{13.0}$ & $0.5$ & $11$ & $11$ &this work \\
         PonosV & $1.2 \cdot 10^{13.0}$ & $0.7$ & $6$ & $5$ &\cite{2016ApJ...824..144F} \\
         PonosQ & $6.5 \cdot 10^{12.0}$ & $0.7$ & $6$ & $6$  &\cite{2016ApJ...824..144F} \\
         Milky Way Satellites Prior & $1.7 \cdot 10^{13.0}$ & $0.0$ & $1-2$  & $4-8$  & \cite{nadler2021dark} \\
         Symphony & $1.0 \cdot 10^{13.0}$ & $0.5$ & $6$ & $6$  &\cite{nadler2023symphony} \\
         \hline
    \end{tabular}
    \caption{
        Amplitude of the evolved PSHMF, given in terms of the $f_{\mathrm{s}} \cdot \Sigma_{\mathrm{sub}}$ (see \refeq{eq_massfunction_proj_evolved}).
        In the ``scaled'' column, the number density has been extrapolated to a $10^{13} \mathrm{M}_{\odot}$ halo at redshift $z=0.5$ using the scaling relations presented in this work. 
        A logarithmic slope of the subhalo mass function $\alpha = -1.93$ was assumed.
    }
    \label{tab_fits_external}
\end{table*}

\begin{table*}
 \begin{tabular}{lcc}
  \hline
    Model & $k_1$ (mass scaling) & $k_2$ (redshift scaling)\\
  \hline
        \gal (unevolved)                                             & 0.55 & 0.37   \\
    host density profile                                             & 0.50 & 0.30   \\
  \hline
        \gal (evolved)                                               & 0.37 & 1.05   \\
        \gal with central galaxy (evolved)                           & 0.43 & 1.18   \\  
        \hanetal ($\gamma = 0.8$)                                    & 0.37 & 0.53   \\
        \hanetal ($\gamma = 1.0$)                                    & 0.35 & 0.61   \\
        \hanetal ($\gamma = 2.0$)                                    & 0.30 & 0.87   \\
        \hanetal ($\gamma$ Interp)                                   & 0.23 & 0.62   \\
    \hline
 \end{tabular}
 \caption{
            The best fits to power law scaling relations predicted by our \gal models and estimates using the \hanetal model.
            The upper section gives the scaling for the unevolved case, while the lower section gives the scaling results for the evolved case.
            We compare the scaling relation for the unevolved case for the scaling relation which would be obtained if the scaling was entirely determined by the host's density profile evolution (\refeq{eq_han_unevolved}).
            For the evolved case, we compare the PSHMF scaling prediction to that of the \hanetal model (\refeq{eq_han_evolved}).
            The transfer function from the host's smooth dark matter density profile is parameterized as a power law, with logarithmic slope $\gamma$ (\refeq{eq_han_transfer}).
            We consider two possibilities for $\gamma$, one where $\gamma$ is a constant and the other with $\gamma$ a function of halo mass.
            For the case of mass independent $\gamma$, we compute the scaling coefficients for a range of physically meaningful values of $\gamma$.
            Finally, to study the effect of $\gamma$ varying systematically with mass, we linearly interpolate the best-fit $\gamma$ values as a function of mass (\reftable{tab_han_fits}) to compute the scaling relations (``$\gamma$ interp'').
           }
  \label{tab_scaling}
\end{table*}

\section{Discussion}\label{sec_discussion}

\subsection{Subhalo Populations}\label{sec_discuss_pop}
In this work, we present updated predictions for the projected number density of subhalos as a function of halo mass and redshift.
The \gal predictions for the unevolved SHMF agree with findings of Van Den Bosch et al. \citep{10.1111/j.1365-2966.2005.08964.x}, that the unevolved SHMF can be written in a universal form (\refeq{eq_scaling_unevolved}). 
Furthermore, we find that \gal agrees with the empirical formula of Gao et al. \citep{10.1111/j.1365-2966.2004.08360.x}, that for the unevolved case the SHMF scales as a function of host halo mass and the abundance of halos in the universe as a whole (\refeq{eq_scaling_evolved}). 
Comparing the \gal  and \symphony results, the predictions for the normalization of the subhalo mass function within the inner $50$ kpc, we find that \gal predicts on average nearly twice the subhalos in this region.
Similarly, Over the entire virial volume we find similar results, with \symphony predicting around $40\%$ the number of subhalos as \galns.
We note that \symphony and \gal agree within halo to halo scatter, and N-body simulations such as \symphony and \cat disagree on the $25\%$ level \citep{nadler2023symphony}.
We also find that the suppression of the subhalo mass function due to the potential of the central galaxy is negligible when compared to the current theoretical uncertainty. 
Similarly, we find that \hanetal \citep{10.1093/mnras/stv2900} is in reasonable, although not perfect, agreement with \gal on the profile of the spatial distribution as well as the halo mass and redshift scaling in projection. 
Here, we provide further discussion for each of these results.

Here, we discuss the expectations for the spatial distributions from the \hanetal \citep{10.1093/mnras/stv2900} model, and how they compare to predictions from \galns. 
The \hanetal \citep{10.1093/mnras/stv2900} model describes the unevolved spatial distribution of subhalos as following the dark matter density distribution of the host halo.
As shown in \reffig{fig_spatial_ratio}, our \gal models show a difference between the unevolved subhalo spatial distribution and the host's smooth dark matter density profile (normalized to $1$ at $r = r_{\mathrm{v}}$).
This is difficult to explain with dynamical friction, as the dynamical friction time scale is proportional to the ratio of the subhalos mass to the host halo's mass ($m / M_\mathrm{h}$) and is longer than age of the universe for sub-halos with with less than $1 / 20$ the mass of the host halo \citep{boylan2008dynamical}.
However, the host halo in our models is evolving with time and older subhalos may have fallen in when the host halo was much less massive (therefore $m / M_\mathrm{h}$ is much closer to $1$), therefore the dynamical friction timescale would be much smaller for these older subhalos.
To determine if dynamical friction is the cause of this difference, \gal could be ran with and without dynamical friction enabled. 
We leave this to future work.
The \hanetal \citep{10.1093/mnras/stv2900} model does not include effects of dynamical friction, while the \gal models does.
For the evolved case, \hanetal \citep{10.1093/mnras/stv2900} use a power law profile to model the ratio of evolved spatial distribution to the host's dark matter density profile. 
\reffig{fig_spatial_ratio} shows that \gal does not follow this expectation exactly, with deviations from a power law near the virial radius.

When comparing our \gal models to Symphony, we find that \gal predicts a normalization of the projected SHMF a factor of $\sim 2$ times higher than Symphony.
We note that this difference remains within $1 \sigma$ halo to halo scatter and caution should be taken when interpreting this result due to the small number of subhalos in this volume which makes obtaining good statistics challenging.
If the difference is statistically significant, there are several possible explanations for the discrepancy.
A large body of works has drawn attention to artificial disruption present in N-body simulations \citep{10.1093/mnras/stx2956,2022MNRAS.517.1398B,10.1093/mnras/stz2767,10.1093/mnras/stz3349}.
An estimate for spurious suppression due to artificial disruption is $10\text{--}20\%$ over the entire virial volume, only increasing to a maximum factor of $3$ within the inner $2 \%$ of the virial radius \citep{2022MNRAS.517.1398B}.
Another possibility is inaccuracies in halo finding algorithms.
We analyze the \symphony subhalo catalogs extracted using the Rockstar \citep{behroozi2012rockstar} halo finder.
Recent work by Mansfield et al. \citep{mansfield2024symfind} has identified $15\text{--}40 \%$ more subhalos within the virial radius when compared to the Rockstar results, increasing to $35\text{--}120\%$ within $r < r_{\mathrm{v}} / 4$.
Another possibility may be the \gal physics models.
Our \gal model does not include a model of pre-infall tidal stripping \citep{behroozi2014mergers}.
We do not make a determination if the discrepancy is numerical in nature and leave determining of the underlying cause(s) of the differences to future work.

Next, we discuss the scaling the projected SHMF as a function of halo mass and redshift.
For the evolved case, our scaling relations from \gal and our analytic models predict that for a factor of $10$ increase in host halo mass the evolved PSHMF function will increase by a factor $\approx 2$.
This is similar to the factor of $\approx 3$ used in the extrapolation performed in Xu et al. \citep{10.1093/mnras/stu2673}, but differs from the factor of $\approx 8$ used in Gilman et al. \citep{10.1093/mnras/stz3480}.
Differences from Gilman et al. \citep{10.1093/mnras/stz3480} are likely due to a change in \galns, in particular an update to the treatment of higher order substructure.
The change in higher order substructure treatment had a large impact on the spatial distributions, with the previous spatial distributions of subhalos being more cuspy. 

Finally, we compare our scaling results from \gal to the \hanetal \citep{10.1093/mnras/stv2900} model of the spatial transfer functions.
Three forms of the tidal stripping transfer function, $\hat{T}$ are considered (see Equations \ref{eq_han_evolved} and \ref{eq_scaling_evolved}), with fits to the projected scaling relation fits provided in table \ref{tab_scaling}.
All cases considered result in host halo mass scaling similar to \galns, with a $10$ times increase in mass resulting in a roughly $\approx$ 2 times increase in the normalization of the PSHMF.
Fixing $\hat{T}(r / r_{\mathrm{v}})$ (see equations \ref{eq_han_evolved} and \ref{eq_han_transfer}) to be invariant in halo mass ($\gamma = 0.8, 1.0, 2.0$ and ``\gal interp'' cases in \reftable{tab_scaling}) results in a best fit of $k_1 \approx 0.3$.
Allowing $\gamma$ to vary with halo mass (linearly interpolating in between $\gamma = 0.94$ and $\gamma = 1.23$) results in $k_1 = 0.23$, corresponding to a $1.7$ times increase in normalization PSHMF with a $10$ times increase in halo mass.
Redshift scaling results ($k_2$) vary from $0.61$ to $0.87$, with no models tested matching the predictions from \gal exactly.
A possible reason for these discrepancies is the difference in spatial distribution between \gal and the \hanetal \citep{10.1093/mnras/stv2900} model. 
Additionally, there may be evolution of the spatial distribution as a function of halo mass and redshift that is not captured in the \hanetal \citep{10.1093/mnras/stv2900} models.

\subsection{$\Sigma_{\mathrm{sub}}$ Prior}\label{sec_discuss_sigsub_prior}
Here, we discuss our recommendations for future priors on the number density of subhalos within the Einstein radius.
To ensure matching of observable quantities, we use our \gal models to recommend a prior on the amplitude of the evolved halo mass function $f_{\mathrm{s}} \cdot \Sigma_{\mathrm{sub}}$, instead of $\Sigma_{\mathrm{sub}}$ directly. 
Our \gal models predict an average number density of $f_{\mathrm{s}} \cdot \Sigma_{\mathrm{sub}} \approx 10^{-2} \text{~kpc}^{-2}$, while the Symphony suite predicts $f_{\mathrm{s}} \cdot \Sigma_{sub} \approx 5 \cdot 10^{-3} \text{~kpc}^{-2}$ (additional results are given in \reftable{tab_fits_external}).
Here, we do not investigate the possible reasons for the discrepancies, and instead treat each results as equally likely to estimate a theoretical uncertainty.
Across works compared here, we find that the value of \fssigsub~ varies from $10^{-3} \text{--} 10^{-2} \text{~kpc}^{-2}$, spanning approximately an order of magnitude, with more recent works in the $5 \cdot 10^{-3} \text{--} 10^{-2} \text{kpc}^{-2}$ range.
In a $100 \text{~kpc}^2$ aperture, this approximately corresponds to a range of $0.1$ to $1$ subhalos with bound mass $8 < \log_{10} (m / \mathrm{M}_\odot)  \leq 9$ (or equivalently a number density of $10^{-3} \text{--} 10^{-2} ~\text{kpc}^2$ in this mass range).
For reference, an average Einstein radius of $1 \text{~arcsecond}$ at redshift $z=0.5$ corresponds to a $\sim 150 \text{~kpc}^2$ aperture. 
We note that this is well within the allotted uncertainty of previous lensing studies.
Furthermore, we estimate that the impact of a central galaxy (reduction in $f_{\mathrm{s}} \cdot \Sigma_{\mathrm{sub}}$ by a factor of $\sim 10\%$) is negligible compared to the current theoretical uncertainty between different models.

We recommend a prior centered around $f_s \cdot \Sigma_{\mathrm{sub}} = 10^{-2} ~\text{kpc}^{-2}$ and allowing for at least a factor of $2$ ($\sim 0.3$ dex) uncertainty.
We note that this is a similar prior to the tighter prior considered by Gilman et al. \citep{gilman2024turbocharging}.
When Gilman et al. \citep{gilman2024turbocharging} compared a  1.5 dex prior uniform prior (similar to that used in previous lensing studies) with  a tighter 0.2 dex Gaussian prior, the lower bound on the half-mode-mass improved by 2 dex.
By implementing a 0.3 dex prior we expect similar improvements on constraints on half mode mass.

\section{Summary}\label{sec_conclusion}
In this work we study predictions for the population level statistics of subhalos of group mass halos at small projected distances.
Particular focus is placed on the normalization of the projected SHMF, where we give recommendations for a more informed prior on this quantity than previous lensing studies. 
Here we provide a summary of key results in our paper:
\begin{itemize}
    \item
    We present a new suite of \gal models of the substructure of host halos with masses $12 \leq \log_{10}(M_{h}/\mathrm{M}_{\odot}) \leq 13.5$ and redshifts $0.2 \leq z \leq 0.8$.
    \item
    Using these simulations, we measure the projected number density of subhalos near the Einstein radius.
    The scaling in halo mass and redshift is well described with a power law with coefficients $k_{1}=0.55$ (halo mass) and $k_{2}=0.37$ (redshift) for the unevolved case and $k_{1}=0.37$ and $k_{2}=1.05$ for the evolved case.
    \item
    The evolved and unevolved projected SHMF amplitude scales differently as a function of halo mass and redshift due to different spatial distributions between the two cases. 
    For the evolved case, the projected SHMF amplitude approximately scales with the cube root of the host halos mass and nearly linearly in redshift.
    We find the scaling of the PSHMF in redshift and mass to be in excellent agreement match between simulations and analytic models.
    \item
    On the group scale, a central galaxy reduces the normalization of the evolved SHMF by $< 5\%$ over the entire virial volume, increasing to $15 \%$ for the PSHMF within the inner $20$ kpc.
    This is much less than the current theoretical uncertainty between different models/N-body simulations.
    \item
    We find that all models/simulations considered agree on the projected SHMF normalization well within a factor of $\sim 2$.
    Additionally, we find that \gal and \symphony agree within the $1 \sigma$ halo to halo scatter for subhalos halos in the bound  mass range $9 < \log_{10}(m_b/\mathrm{M}_\odot) \leq 10$.
    Current lensing studies adopt sufficiently wide priors (of nearly a factor of $30$) to account for theoretical uncertainties in $f_{\mathrm{s}} \cdot \Sigma_{\mathrm{sub}}$.
    We recommend using a stronger prior on \fssigsub when compared to previous lensing studies, centered around $f_s \cdot \Sigma_{\mathrm{sub}} = 10^{-2} \text{kpc}^{-2}$ and allowing for at least a factor of $2$ uncertainty.
    Compared to previous lensing studies, we can expect up to a $2$ dex improvement when using this tighter prior.

\end{itemize}

\begin{acknowledgments}
We thank Ethan Nadler and Risa Wechsler for thoughtful comments and suggestions.

This research was conducted using Pinnacles (NSF MRI, \# 2019144) at the Cyberinfrastructure and Research Technologies (CIRT) at University of California, Merced.

This research was supported in part by grant NSF PHY-2309135 to the Kavli Institute for Theoretical Physics (KITP).

AN and CG acknowledge support from the NSF through AST-
2206315 “Collaborative Research: Measuring the physical proper-
ties of DM with strong gravitational lensing” and through JWST-GO program \#2046 which was provided by NASA
through a grant from the Space Telescope Science Institute, which is
operated by the Association of Universities for Research in Astronomy, Inc., under NASA contract NAS 5-03127.

DG acknowledges support for this work provided by the Brinson Foundation through a Brinson Prize Fellowship grant, and from the Schmidt Futures organization through a Schmidt AI in Science Fellowship.

\end{acknowledgments}

\appendix

\section{Evolution of the Central Galaxy}\label{sec:appendix}
\begin{figure*}
    \includegraphics[width=\textwidth]{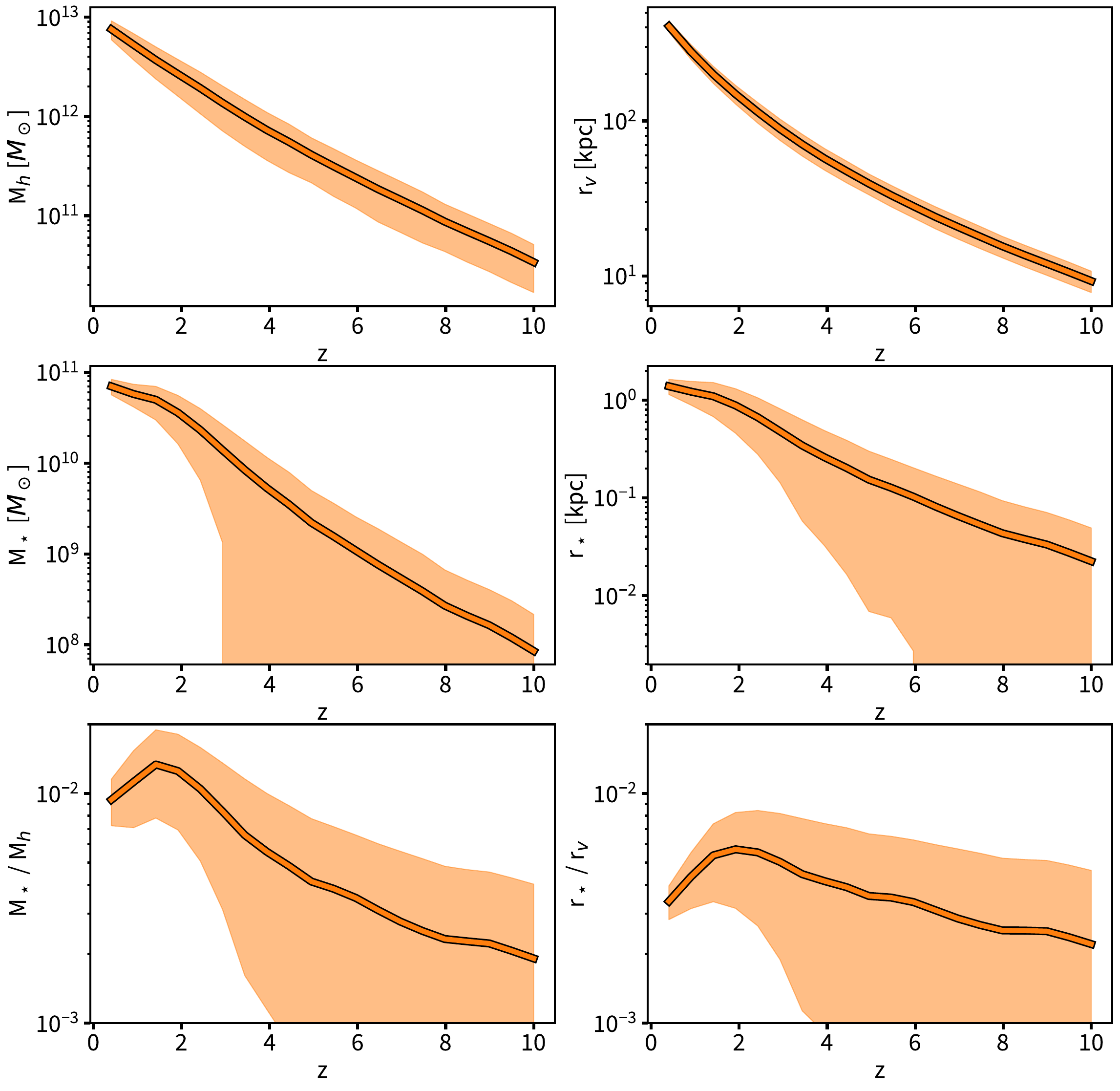}
    \caption{
               Our \gal model predictions for the evolution of a $10^{13} \mathrm{M}_\odot$ halo and its central galaxy.
               Stellar masses are determined using the Behroozi et al. relation between halo mass and stellar mass, while galactic radii are determined using empirical fits to the stellar mass to stellar radius relation provided by Shen et al.
               For simplicity, no scatter has been included in either of these relationships.
               The top left panel gives the evolution of the mass of the central halo while the top right panel gives the evolution of the virial radius of the halo.
               The middle left panel shows the evolution of the stellar mass, while the middle right panel shows the evolution of the Hernquist radius of the central galaxy.  
               Finally, the lower left and right panels show the evolution of the ratio of stellar mass to halo mass and Hernquist radius to virial radius, respectively.
               The curve shows the mean value, while the shaded region shows $1 \sigma$ scatter. 
     }
    \label{fig_um_evolution}
\end{figure*}

We plot properties of the central galaxy in our \gal~ models.
\reffig{fig_um_evolution} shows the evolution of the central galaxy's mass and radius over redshift.

\bibliography{gannon2024}

\end{document}